\documentclass[12pt]{article}
\input epsf.sty

\newcommand{\sect}[1]{\setcounter{equation}{0}\section{#1}}

\newcommand{\eqn}[1]{(\ref{#1})}
\newcommand{\br}{\nonumber \\}
\newcommand{\be}{\begin{equation}}
\newcommand{\ee}{\end{equation}}
\newcommand{\bea}{\begin{eqnarray}}
\newcommand{\eea}{\end{eqnarray}}
\newcommand{\p}{\partial}

\newcommand{\ov}{\over}

\newcommand{\plb}[3]{Phys. Lett. {\bf B#1} ({#2}) {#3}}
\newcommand{\prd}[3]{Phys. Rev. {\bf D#1} ({#2}) {#3}}
\newcommand{\hepth}[1]{{\tt hep-th/{#1}}}
\newcommand{\hepph}[1]{{\tt hep-ph/{#1}}}

\begin{document}
\renewcommand{\thefootnote}{\fnsymbol{footnote}}
\begin{titlepage}
\begin{flushright}
SINP/TNP/2007/23
%hep-th/\\
\end{flushright}
\vskip .7in
\begin{center}
{\Large \bf Assisted Inflation from Geometric Tachyon} \vskip .7in
{\large Kamal L. Panigrahi $^a$\footnote{e-mail: {\tt
panigrahi@iitg.ernet.in}} and \large Harvendra Singh
$^b$\footnote{e-mail: \tt h.singh@saha.ac.in}} \vskip .2in
{$^a$}{\it Department of Physics\\
Indian Institute of Technology, Guwahati\\
781 039, Guwahati, India} \vskip .2in
{$^b$}{\it Theory Division, Saha Institute of Nuclear Physics\\
1/AF Bidhannagar, Kolkata 700064, India} \vspace{.7in}
\begin{abstract}
\vskip .5in \noindent We study the effect of rolling of $N$
D3-branes in the vicinity of NS5-branes. We find out that this
system coupled with the four dimensional gravity gives the slow
roll assisted inflation of the scalar field theory. Once again
this expectation is exactly similar to that of $N$-tachyon
assisted inflation on unstable D-branes.
\end{abstract}
\end{center}
\vfill

\end{titlepage}
\setcounter{footnote}{0}
\renewcommand{\thefootnote}{\arabic{footnote}}
\tableofcontents \sect{Introduction}

%%%%%%%%%%%%%%%%%%%%%%%%%%%%%%%%%%%%%%%%%%%%%%
During  inflation \cite{guth} the spacetime must expand with an
accelerated expansion. Allowing this phenomenon in the early
universe solves many cosmological puzzles; e.g. the isotropy and
homogeneity of the large scale structures and the flatness of the
universe. Inflation is also considered to be the most plausible
source of the primordial density perturbations in the universe. The
conventional inflationary models deal with a scalar field rolling
down  a slowly varying potential in FRW spacetime \cite{linde}. But,
it is now well accepted that string theory also has many
cosmological vacua which have inflationary  properties, {\it a la}
\cite{kklt,kklmmt}. In fact there are plentiful of them. The early
models, called brane-inflation models \cite{dvali}, have an
attractive brane-antibrane force which drives the modulus (inflaton)
separating the branes. But, generally, the inflation is not the
slow-roll one there. The  brane-antibrane models in a warped
compactification scenario \cite{kklmmt}, partly solve the slow-roll
problem, but lack the precise knowledge of the inflaton potential in
presence of fluxes. There are other models based on open-string
tachyon condensation which are capable of providing inflation, see
\cite{anup,sen3,hs1,sinha,hs2,zamarias,anne,cai,hsingh}. In general,
the tachyon models are also plagued with the same large
$\eta$-problem as the conventional models and are not favoured for
slow-roll inflation, see \cite{kofman}. Although, this difficulty
can be overcome by allowing a large number of tachyons to
simultaneously roll down and assist the inflation \cite{hsingh}.
There are  number of various other attempts to provide stringy
inflation models. These are for instance inflationary models based
on branes intersecting at special angles \cite{garcia}, and also the
D3/D7 models  where the distance modulus plays the role of inflaton
field \cite{keshav,chen}. The race-track inflation models driven by
closed string modulus could be found in \cite{racetreck}.

Time dependent solutions in string theory are very interesting and
challenging in view of solving puzzles in early universe
cosmology. However there are quite a few number of known examples
of such solutions in string theory/supergravity. In particular the
rolling tachyon solution\cite{sen1} represents the decay of the
unstable branes by the rolling of the open string tachyon in the
potential valley. The end product of such a rolling process has
been shown to comprise of a closed string excitation--the tachyon
matter. Various cosmological applications has also been studied
using this open string tachyon solution of string theory. Not long
back Kutasov \cite{kutasov1} proposed an example of such a time
dependent solution in string theory, whose properties are
strikingly similar to that of rolling tachyon solution of open
string models. This is the D-NS5 brane system, where the dynamics
of the BPS D-brane has been studied in the vicinity of a stack of
BPS NS5-branes. As such this system is nonsupersymmetric as
D-brane and NS5-branes break different halves of supersymmetry.
The time-dependent dynamics has been studied both from the point
of view of effective field theory and from the full string theory
by constructing the relevant boundary states in NS5-brane near
horizon geometry. It has been shown that when the Dp-branes are
very far away from the NS5-branes there is the usual gravitational
interaction between the two, and the D-branes will start to move
towards the NS5-brane. When the D-brane is close enough to the
NS5-brane (at a distance below $l_s$) it behaves exactly like that
of the rolling tachyon. It has also been shown that there are no
open string degrees of freedom at this point and all the other
qualitative behaviour of the open string tachyons are also
present. Hence the rolling tachyon of the open string models get a
geometrical meaning- it is the `radial' distance between the
D-brane and the NS5-branes. This tachyon-radion correspondence has
been analyzed more from the full string theory view point
\cite{NST-BS} and considerable amount of work has been done in
connection with this (see for example \cite{Rbrane} and more
recently \cite{NS5}). The cosmological applications of this
solution has been analyzed in detail in the literature
\cite{NS-cosm}. The inflationary scenario has also been worked out
when the NS5-branes are distributed over a particular ring in the
transverse space. It has been shown that the slow roll conditions
and number of e-foldings are consistent with that of a
4-dimensional model with a perturbative string coupling. The
traditional reheating mechanism has also been reproduced in this
set up.

In \cite{cai,hsingh} $N$-tachyon assisted inflation was studied,
where all the $N$ tachyon fields on the non-BPS branes have been
allowed to roll simultaneously in the valley of the potential.
Coupling this system of $N$ unstable D3-branes to gravity the
inflationary scenario has been observed and the slow roll
parameters are shown to be consistent with the assisted inflation
of the $N$ scalar fields.

Knowing the results of the tachyon assisted inflation on a system
of $N$ non-BPS branes and the by now well known correspondence
between the tachyon-radion, it is tempting to analyze the assisted
inflation in the (D-NS5)-brane system. The geometric tachyon can
be viewed as the field responsible for inflation, and one could
study the slow roll parameters, e-foldings and critical number
density. We have addressed this problem in the present paper. We
assume that all the $N$ number of D3-branes roll simultaneously
into the NS5-brane throat. All the D3-branes are coincident, hence
the stack of parallel and coincident branes is BPS. We further
assume that all the perturbative open string tachyons between the
D3-branes are switched off. Hence the dynamics of this system is
governed by the so called geometric tachyon.

Rest of the paper is organized as follows. In section-2, we review
the basic ideas of assisted inflation. Section-3 is devoted to the
study of assisted inflation from geometric tachyon, by studying
the system of D-branes into the NS5-brane throat. In section-4, we
make some numerical studies. We conclude in section-5 with some
comments.

%\sect{Tachyon Assisted Inflation}

\section{Brief Review: Assisted Inflation}

The assisted inflation idea was proposed in \cite{liddle} to overcome the
large
$\eta$ problem in scalar field driven inflaton models. We review the main
aspects of that work here. We consider a scalar field with
potential $V(\phi)$  minimally coupled to Einstein gravity \cite{linde}
\be
\int d^4x \sqrt{-g}[{M_p^2\over 2} R -{1\over 2}(\partial\phi)^2
-V(\phi)]
\ee
where four-dimensional Planck mass $M_p$ is related to the Newton's
constant $G$ as
$M_p^{-2}={8\pi G}$. Considering purely time-dependent  field
in a spatially flat FRW spacetime
\be
ds^2=-dt^2 +a(t)^2\left({dr^2}+r^2(d\theta^2+sin^2\theta
d\phi^2)\right)
\ee
the classical field equations can be written as
\bea
&& \ddot\phi=-V,_{\phi}-3H\dot\phi \br
&&H^2={8\pi G\over 3}({\dot\phi^2\ov2}+V)
\label{dgt1}
\eea
where $H(t)\equiv\dot a/a$ is the Hubble parameter. This simple model has
proved to be a prototype for explaining the mechanism of
inflation in early universe. For example, if we take a quadratic
potential $V=m^2\phi^2/2$ and let the field roll down from some large
initial value, the field will roll down to its minimum value $\phi=0$ and
 so spacetime will inflate \cite{linde}. But the
inflation has to be a slowly
rolling one in order to fit with cosmological observations. That is
the field $\phi$ must vary slowly such that
there is a vanishing acceleration,  $ \ddot\phi\sim0$. Under the slow-roll
conditions,
the time variation of $\phi$  gets
related to the slop of the potential as
$$\dot\phi\simeq-{V,_{\phi}\over
3 H}.$$
So the potentials with gentle slop are preferred for a good
inflation.

The standard slow-roll parameters are \cite{sasaki}
\be
\epsilon\equiv{ M_p^2\over2} {V'^2\over V^2},~~
\eta\equiv{ M_p^2} {V''\over V}
\ee
where primes are the derivatives  with respect to $\phi$. For the
slow-roll inflation
both $\epsilon$ and $\eta$ will have to be small. Also these parameters
are in
turn related to the spectral index, $n_s$, of the scalar density fluctuations
in the early universe as
\be n_s-1\simeq -6\epsilon+2\eta.
\ee
A nearly uniform power spectrum observed over a wide range of
frequencies in the density
perturbations in CMBR measurements \cite{wmap}, however, requires
$n_s\simeq .95$. It can
be achieved only if $$\epsilon\ll1,~~~\eta\ll 1.$$
These are some of the stringent bounds from cosmology which inflationary
models have to
comply with.

For single scalar field the power spectrum of the scalar curvature
perturbations can be written as \cite{sasaki}
\be
P_R={1\over 12\pi^2M_p^6}{V^3 \over V'^2}
\ee
The  amplitude (or size) of the density fluctuations are governed by
\be\label{df1}
\delta_s={2\over5}\sqrt{P_R}={1\over 5\sqrt{3}\pi M_p^3}{V^{3/2} \over
V'} \le 2\times 10^{-5}
\ee
The inequality on the right side of equation \eqn{df1} indicates the
COBE bounds on the size of the density
perturbations at the beginning of the last 50 e-folds of
inflation.\footnote{ The number of e-folds, $N_e$, during inflationary
time interval $(t_f-t_i)$ are estimated as $N_e=\int_{t_i}^{t_f} H dt$.
The
universe requires 50-60 e-folds of expansion in order to explain the
present size of the observed large scale structure.}

It can be easily seen that
 the models with quadratic potentials are not useful for
inflation as they are plagued with so called large $\eta$-problem. From
the above we find that $\eta\sim 2 M_p^2/\langle\phi\rangle^2$. Therefore
 $\eta$ can be small only if $\phi$ has trans-Planckian vacuum
expectation value during inflation. But
allowing the quantum fields to have
trans-Planckian vev will spoil the classical analysis and will involve
quantum corrections to the potential.

An effective resolution of the large $\eta$ problem can come
from assisted inflation idea \cite{liddle}. The model  involves
large number of scalar
fields $\phi_i$ $(i=1,2,\cdots,N)$. Let us demonstrate it here for a
 quadratic potential $V=m^2\phi^2/2$ where all scalars
have the same mass $m$. The scalar fields are taken to be
noninteracting but are minimally coupled to gravity. The equations
of motion are \bea &&\ddot \phi_i= -{m^2 \phi_i}-3 H \dot \phi_i
~~~~{\rm for~each } ~i\ ,\br && H^2= {8\pi G \over 3 }\sum_{i=1}^N
(V(\phi)+ {\dot \phi_i^2\over2})\ . \label{aeqn1} \eea Now, if all
the fields are taken to be identical
$\phi_1=\phi_2=\cdots=\phi_N=\Phi(t)$, the simplified equations
become \bea &&\ddot \Phi= -{m^2 \Phi}-3 H \dot \Phi\br && H^2= {8\pi
G \over 3 } N (V(\Phi)+ {\dot \Phi^2\over2}) \label{aeqn2} \eea One
finds that due to $N$ scalar fields, the eqs.\eqn{aeqn2} are having
an effective Newton's constant  $G.N$ when compared with the
eqs.\eqn{dgt1}. So we easily determine that $$\epsilon=\eta={2
M_p^2\over N \Phi^2} \ .$$ Thus for sufficiently large $N$,
$<\Phi>$ need not be trans-Planckian. It can also be seen that since
$H\sim {\cal O}(\sqrt{N})$ the spectral index
 \be
n_s-1=2\eta-6\epsilon\simeq 2{\dot H\over H^2}\sim {\cal O}({1\over N})\
.
\ee
If larger is the value of $N$, the index $n_s\sim 1$ which refers to the
flatness (or no-tilt) of
the spectrum of the density fluctuations in the observed universe.
 However recently in \cite{Baumann} an alternative way of
warped compactification has been used to control the large $\eta$
problem.

\section{Inflation from $N$-Geometric Tachyon}
In this section, we will study the inflation from the motion of
$N$ D3-branes falling into the NS5-brane throat geometry. The
metric $(g_{\mu\nu})$, dilaton $(\phi)$ and Neveu Schwarz field
$H_{mnp}$ of a system of $k$ coincident $NS5$-branes is given by:
\bea ds^2 &=& \eta_{\mu \nu} dx^{\mu} dx^{\nu} + H
(x^n)\delta_{mn}dx^m dx^n \cr & \cr e^{2(\phi-\phi_0)} &=& H(x^n),
\>\>\> H_{mnp} = \epsilon^q_{mnp} \p_{q} \phi, \>\>\> H (r) = 1 +
\frac{k \alpha'}{r^2} ,\eea where $r = |\vec{x}|$ is the radial
coordinate away from the NS5-branes in the transverse space $R^4$.
The effective action on the world volume of $Dp$-brane is governed
by the DBI action:
\begin{eqnarray}
S_{p} = -T_{p} \int d^{p+1} \xi e^{-(\phi
-\phi_0)}\sqrt{-\det(G_{ab} + B_{ab})}.
\end{eqnarray}
Where $G_{ab}$ and $B_{ab}$ are the induced metric and the
B-field, respectively, onto the world volume of the $Dp$-brane:
\begin{eqnarray}
G_{a b} &=& {{\partial X^{\mu}}\over{\partial \xi^a}} {{\partial
X^{\nu}}\over{\partial \xi^b}}g_{\mu\nu}, \cr & \cr B_{a b} &=&
{{\partial X^{\mu}}\over{\partial \xi^a}} {{\partial
X^{\nu}}\over{\partial \xi^b}}B_{\mu\nu},
\end{eqnarray}
where $\mu$ and $\nu$ runs over whole ten dimensional space time.
The worldvolume coordinates of $Dp$-brane are leveled by
$\xi^{a}$ $(a = 0,...,p)$, and we set (by reparameterization
invariance on the world-volume of the $Dp$-brane) $\xi^a = x^a$.
The DBI action is given by \bea S_p = -\tau_{p}\int d^{p + 1}~x
\frac{1}{\sqrt{H}} \sqrt{1 + H \p_a R \p^a R} \eea Compare with
the tachyon effective action of the open string models \bea
{\mathcal S}_{tach} = - \int d^{p + 1}~x V(T) \sqrt{1 + \p_{a} T
\p^a T} \label{act-tach} \eea Comparing the above two actions, one
gets

\bea  \pm\frac{dT}{dR} &=& \sqrt{1 + \frac{k\alpha'}{R^2}} \cr &
\cr V(T) &=& \frac{\tau_p}{\sqrt{H(R(T))}}\eea

The differential equation above, has the solution
$$ \pm T= \sqrt{R^2 + a^2} -a \log ({a + \sqrt{R^2 + a^2}\over R}) $$

where $ a^2 = k \alpha'$.  For the minus sign the  potential $V$
has a universal minimum at $T=\infty$ or $R=0$ while it has a
maximum at $T=-\infty$ or $R=\infty$.  At $t=0$ , the tachyon will
roll down from $T=-\infty$ to its minimum at $T=\infty$ at some
later time. For the upper sign it will be opposite of this. We
must employ the complete potential for the cosmological study
although for the inflation purpose it will happen only near the
top of the potential at $T=-\infty$. At the top the potential has
a behaviour ($T^2\gg a^2$) roughly as

$$   V(T)\sim \tau_p (1- {a^2\over 2T^2}) +O({a^4\over T^4})$$

Our aim is to study the four dimensional cosmology of this system.
For that purpose, we would like to solve the equations of motion
derived from the effective action of the `geometric tachyon' with
the above potential for the D3-brane falling into the NS5-brane
throat. In what follows we only consider the homogeneous mode, where
the transverse coordinates of the NS5-branes depend on time
coordinate only. To make it look familiar with that of the open
string tachyon effective action on unstable D-branes to the
quadratic order, we make a simple rescaling of the tachyon field
$T\rightarrow \sqrt{\alpha'}T$.  After the rescaling the effective
action and the potential for the tachyon can be written as
\footnote{Though we are taking large number of D3-branes, the
NS5-branes are much heavier (of the order $\frac{k}{g^2_s}$) than
the Dp-branes (of the order $\frac{N}{g_s}$) in the weak string
coupling regime. For  sufficiently weak coupling, the back reaction
of these probe branes can be ignored. Further we restrict ourselves
to the non-interacting geometric tachyon modes only in the DBI action,
which are the radial distances from the core of the NS5-branes. We are
ignoring other excitations on the world volume of D3-branes including
gauge fields. Hence in the
lowest order analysis only geometric tachyons will contribute.}

\bea S = -\sum^{N}_{i =1} \int d^4 x V_i (T_i)
\sqrt{-\det(g_{\mu\nu} + \alpha' \p_{\mu} T_i (R) \p_{\nu} T_i (R))}
\eea where in the action above, all the potentials have same
functional form, that is \bea V_i(T_i) = V(T_i) = \tau_3 (1- {k\over
2T_i^2}). \eea One can see that the equations of motion for the
purely time dependent tachyon fields are decoupled from each other,
and therefore one can take the factorization ansatz. In other words,
we would like to assume that all the D3-brane roll into the
NS5-brane throat at the same time. We will have the following \bea
T_1 (R(t))&=& T_2 (R(t)) = T_3(R(t)) = \cdots = T_N (R(t)) = \Phi(t)
\cr & \cr V_1(T_1 (R))&=& V_2(T_2 (R)) = V_3(T_3(R)) = \cdots =
V_N(T_N (R)) = V(\Phi(t)) \eea With this ansatz in mind the
effective action for a stack of D3-brane rolling into the NS5-brane
can be given as \bea S = -N \int{d^4 x V(\Phi)\sqrt{-g}\sqrt{1 +
\alpha' (\p_\mu \Phi)^2 }} \eea Now we will couple this system with
that of the four dimensional gravity given by \bea S_{\rm{grav}} =
\frac{M^2_p}{2}\int{d^4 x \sqrt{-g} R} \eea The field equations we
would like to solve combining gravity and the geometrical tachyon
effective action are

\bea \ddot{\Phi}&=& -(1-\alpha' \dot
{\Phi^2})\left(M^2_s \frac{V_{,\Phi}}{V} + 3 H
\dot{\Phi}\right)\cr & \cr H^2 &=& \frac{8\pi G}{3}\left(1 -
\frac{k}{2\Phi^2}\right)\frac{\tau_3 N}{\sqrt{1 - \alpha'
\dot{\Phi^2}}} \eea
where we have taken FRW ansatz.
Assuming $k \ll \Phi^2$ and $\alpha' \dot\Phi^2 \ll 1$, keeping the
leading order terms, we get \bea \ddot{\Phi} &=& \left(\frac{ M^6_s
k}{\Phi^3} - 3H\dot\Phi\right) + \mathcal{O} (\Phi^3) \cr & \cr H^2
&=& \frac{8\pi G \tau_3 N}{3}\left(1 - \frac{k}{2\Phi^2} +
\alpha'\frac{\dot\Phi^2}{2}\right) + \mathcal{O} (\Phi^4) \cr & \cr
&=& \frac{8\pi G}{3}\frac{\tilde N}{g_s} \left(V_{eff} +
\frac{\dot\Phi^2}{2}\right) , \eea where in the last equality \bea
V_{eff} = M^4_s\left(1 -  \frac{k M^2_s}{2\Phi^2}\right), \>\>\>
\tilde N = \frac{N}{(2\pi)^3}, \>\>\> \Phi \to \sqrt{\alpha'}
\Phi\eea

Let us redefine  new fields $\psi=\sqrt{\tilde N\over g_s} \Phi$.
In which case
\bea
H^2 &=& \frac{8\pi G}{3}
\left(V_{eff} + \frac{\dot\psi^2}{2}\right) , \eea
with
\be
 V_{eff}(\psi) = \frac{\tilde N}{g_s}M^4_s\left(1 - \frac{\tilde
N}{g_s} \frac{k M^2_s}{2\psi^2}\right), \>\>\> \ee

\subsection{Slow-roll and the spectrum}
The standard slow-roll parameters can be evaluated by treating $\psi$ as
the
inflaton field. We get
\bea
 \epsilon &=& {M_p^2 \over 2}\left({ V'\over V}\right)^2 \simeq
{  g_s\over 2\tilde  N k } {M_p^2\over M_s^2}\left( {k M_s^2\over
\Phi^2}\right)^3
\nonumber \\
\eta &=& {M_p^2 }{ V''\over V} \simeq
-6 {  g_s\over 2\tilde  N k } {M_p^2\over M_s^2}  \left( {k M_s^2\over
\Phi^2}\right)^2 \ .
\label{slow1}
\eea
 The right most
equalities in the above are expressed in terms of the
geometric tachyon fields $\Phi$. These expressions are valid when
 ${k M_s^2\over  \Phi^2} << 1$.
If we simply take $k=2$ (say), we need to have  vev for fields
$\Phi$ to be greater than the string mass scale! That is the
inflaton fields must have trans-stringy vev's. For this reason it
will be useful to keep $k$, the number of NS5-branes, as small as
possible. These trans-stringy vev's can be realised in this
geometric tachyon model at the place where the D3-branes are far
away from the NS5 branes and start slowly rolling towards
NS-branes. It is obvious that for
$\epsilon, \eta \ll 1$, i.e. for
a slow-roll,  the quantity ${ 3 g_s\over \tilde  N k } {M_p^2\over
M_s^2} \le 1$. Hence we get a bound
 \be
{\tilde N }\ge { 3g_s\over k}{M_p^2\over M_s^2}\ . \label{bound1}
\ee For a large ${M_p^2\over M_s^2}$ ratio and  weak string coupling
the bound can be realised by taking sufficiently large enough $N$.
But $N$ cannot be too large as we will see below.

The next step we will calculate the amplitudes of the primordial density
perturbations.
These amplitudes  can be expressed as
\bea\label{amp1}
\delta_s &=& {1\over \pi\sqrt{75 }}{1\over M_p^3 }{ V^{3/2}\over V'}
\nonumber\\
&&\simeq
{1\over\pi \sqrt{75 k}}
{\tilde  N k\over g_s } {M_s^3\over M_p^3}
\left( { \Phi^2\over k M_s^2}\right)^{3/2}
\label{fgh1}
\eea
The cosmological bounds on the size of these perturbations are
\be
\delta_s \stackrel{<}{\sim}2 \times 10^{-5}
\label{bound2}\ee
which is at the beginning of last 50 e-folds of inflation. If this
bound has to be respected $N$ cannot be very large in \eqn{fgh1}.

Let us take some reasonable data
\be
 k=2, ~ { 3 g_s\over \tilde  N k } {M_p^2\over M_s^2}\sim 1,~{ \Phi^2\over k
M_s^2}\sim 10
\label{data2} \ee
which gives from \eqn{slow1}
\be
\epsilon \sim .002, ~~~~~\eta\sim -.01 \ee
and from the bound \eqn{bound2} on the size of amplitudes we find that the
ratio
$${M_s \over M_p} \le 10^{-5} .$$
That is the string scale has to be  around $10^{14}$ GeV.
However, if we increase $k$ and keeping rest the same, the string scale
can get a higher value. In this later case, geometric tachyons must
acquire even
higher vev's before they roll down. So higher $k$ value is not that
favourable.

The bounds \eqn{bound1}  and \eqn{bound2} immediately tell us
$$ N \sim (2\pi)^3 {3 g_s\over 2}\times 10^{10}$$
which indeed constitutes a large number D3-branes for a given string
coupling.

\section{Numerical analysis}

It is imperative to study the model described earlier numerically.
For this we shall like to study the model directly in terms of the field
$R$
since we know the equations exactly there while in terms of $T$  we
need to make approximations. The
field
equations governing the time evolution can be straight forwardly obtained
\bea
&&\ddot R=- {h'\over 2h} \dot R^2 -(1-h\dot R^2)\left( {V'\over h V}+ 3
H\dot R \right) \cr
&& H^2\equiv{\dot a^2\over a^2}= {8\pi G\over 3} {\tau_3 N V\over
\sqrt{1-h\dot R^2}}
\label{hg1}
\eea
where $h(R)=1 + kl_s^2/R^2$. We shall be studying the situation where
initially  $R^2\gg k l_s^2$ and $R$ rolls down from the flat region of the
potential $V(R)= 1/\sqrt{h}$, also shown in the figure Fig.\eqn{pot1}.
Note that we can also
rewrite the second
equation as
\bea
&& H^2\equiv{\dot a^2\over a^2}= {M_s^2\over 3k} { \tilde  N k \over g_s
} {M_s^2\over M_p^2}
{V\over
\sqrt{1-h\dot R^2}}
\eea
It is a constraint equation.
\begin{figure}[!ht]
\leavevmode
\begin{center}
\epsfysize=5cm
\epsfbox{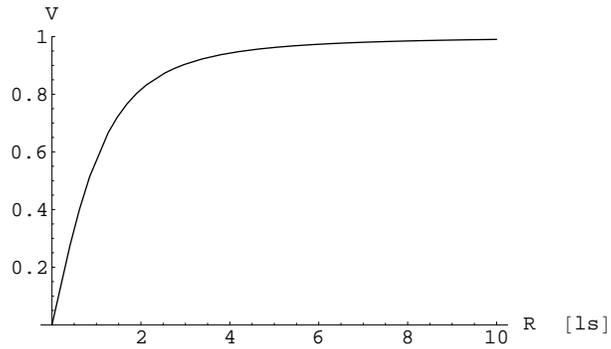}
\end{center}
\caption{\it Plot of $V(R)$ as a function of R for $k=2$. It has a
large flat region and is suitable for eternal inflation.} \label{pot1}
\end{figure}

The two equations can be numerically solved
given some initial conditions.
We shall of course take our standard values from eq.\eqn{data2}, $k=2, {
3 g_s\over 2\tilde  N
k } {M_p^2\over M_s^2}\simeq 1 $ as determined earlier.
We take the initial values at time $t=0$ as
\be R(0)=4.5\,l_s, ~\dot
R(0)=.01
\label{data3} \ee
 and the Hubble
parameter as $ H(0)^2=(1/2) M_s^2$. Note that the time $t$ is
measured in the units of $1/M_s$.
Now solving the equations \eqn{hg1} gives us pretty expected results. The
Hubble
parameter $H$ is plotted in the Fig.\eqn{fig.1}. Initially it remains
almost constant in time describing the slow-roll inflationary epoch.
\begin{figure}[!ht]
\leavevmode
\begin{center}
\epsfysize=5cm
\epsfbox{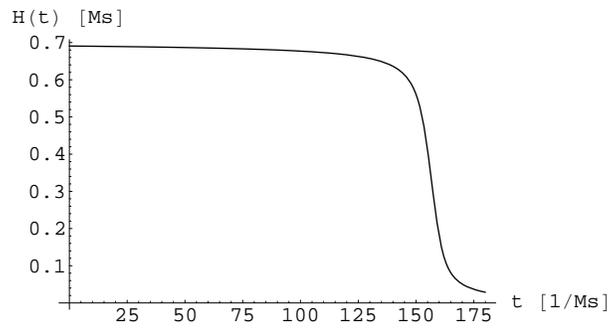}
\end{center}
\caption{\it Plot of $H(t)$ as a function of time. Area under the curve
is roughly 100.} \label{fig.1}
\end{figure}
The area under the H-curve is roughly 100 which gives us an estimate of
the number of e-folds of inflation.
The second graph Fig.\eqn{fig.2} provides the time evolution of the
geometric tachyon $R$.
Thus the initial conditions eq.\eqn{data3} we have chosen do provide us
sufficient
e-foldings before the inflation ends.
\begin{figure}[!ht]
\leavevmode
\begin{center}
\epsfysize=5cm
\epsfbox{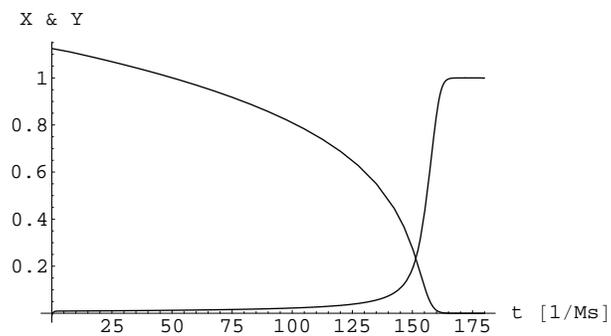}
\end{center}
\caption{\it The quantities $X\equiv R/4$ and $Y\equiv |\sqrt{h}\dot R|$.
}
\label{fig.2} \end{figure}
From figures \eqn{fig.1} and \eqn{fig.2} we find that the
inflation ends where $Y$ becomes of order 1, at the same time $R$
becomes of the order of $l_s$.  This describes the transition
point where inflation ends and the radion matter phase takes over.
This comes out exactly in the similar manner as in open-string
tachyon assisted inflation \cite{hsingh}.

\section{Conclusions}
We have studied in this paper assisted inflation from the motion
of stable D3-branes into the NS5-branes in flat FRW cosmology.
This provides us a further example of assisted inflation in string
theory which involves nonperturbative objects. The slow roll
parameters, when the number of D3-branes are large, has been shown
to be consistent with the expectations from the scalar field
theory observations. The assumption which is self imposed in our
analysis is that the number of probe D3-branes is large. However,
as we have demonstrated that the number of D3-branes also has to
obey certain bound, which comes from the cosmological bound on the
size of small perturbations. Before concluding the present
analysis, we would also like mention what happens in the limit
when the number of NS5-branes are too large and when $N$ is too
small. One can express the size of the cosmological perturbation
as a function of the slow roll parameter, the string scale and the
Planck scale in the following manner. Using eq.\eqn{slow1} during
the slow-roll, one can easily express \eqn{fgh1} as
$$\delta_s= {3 \over \pi\sqrt{75 k}} {1\over|\eta|} {M_s\over M_p}
\sqrt{k M_s^2 \over \Phi^2}.$$ Now for a small value of $N=1$ and
for $k=10^3 M_p^2/M_s^2$, one can show that: $\delta_s =
(.11) (1/|\eta|) \sqrt{1/1000} (M_s/M_p)^2 \sqrt{k M_s^2/
\Phi^2}$. From the observations, we must keep $\eta \sim .01$
during the process of inflation, so $\delta_s \sim .35
(M_s/M_p)^2 \sqrt{k M_s^2/ \Phi^2}$. Thus the amplitudes are very
large unless the ratio $(M_s/M_p)^2 \sim 10^{-5}$. Which implies
$k\sim 10^8$. But the problem with this large $k$ value is that we
have inherently taken $k M_s^2 > M_p^2$, the inflaton field
$\Phi$'s will then be taking trans-Planckian vev's. The lesson is
that we must maintain $k M_s^2 < M_p^2$. So along with eq.
\eqn{bound1} we have a double bound
$$ M_p^2 > k M_s^2 \ge {3g_s\over \tilde N} M_p^2 .$$
Also the ratio of two scales $M_p/M_s$ must follow this bound.
It follows that for large $k$ one has
to keep $N$ sufficiently large, so that the
amplitude of cosmological perturbations is small and the slow roll
parameters could be achieved.

As it is well known that when the D-branes come very close to the
NS5-brane one observes the exponential decaying behaviour of the
potential. It would be interesting to show the assisted inflation
with this exponential potential as well. It would further be
interesting to analyze the assisted inflation with the geometric
tachyon, when the NS5-branes are distributed over a ring. We would
like to return to some of these issues in future. \vskip .2in
\noindent {\bf Acknowledgments:} We would like to thank the
anonymous referee for giving constructive suggestions. KLP would
like to acknowledge the hospitality at the Center for Quantum
Spacetime (CQUeST) at Seoul where a part of this work was done.


\begin{thebibliography}{10}

\bibitem{guth} A. Guth, Phys. Rev. D23 (1981) 347; A.D. Linde,
\plb{108}{1982}{389}; A.R. Liddle and D.H. Lyth, Phys. Rep. 231 (1993) 1;
 A.R. Liddle and D.H. Lyth, {\it Cosmological inflation and
Large Scale Structure}, Cambridge University Press (2000).
E.W. Kolb and M.S. Turner, {\it the Early Universe}, Addision-Wesley
(1990).


\bibitem{linde} A.D. Linde, \plb{129}{1983}{177}.

\bibitem{kklt} S. Kachru, R. Kallosh, A. Linde and S.P. Trivedi,
\prd{68}{2003}{046005}, \hepth{0301240}.

\bibitem{kklmmt}
S. Kachru, R. Kallosh, A. Linde, J. Maldacena, L. McAllister and S.P.
Trivedi, \hepth{0308055}.

\bibitem{dvali} G.R. Dvali and S.H. Tye, \plb{450}{1999}{72},
\hepph{9812483}; C. Burgess, M. Majumdar, D. Nolte, F. Quevedo, G. Rajesh
and R. Zhang, JHEP 0107 (2001) 047, \hepth{0105204};
D. Choudhury, D. Ghoshal, D.P. Jatkar and S. Panda,
JCAP 0307 (2003) 9, \hepth{0305104}.

%\bibitem{mac} R. Easther and L. McAllister, \hepth{0512102}.

\bibitem{anup}
A.~Mazumdar, S.~Panda and A.~Perez-Lorenzana,
   %``Assisted inflation via tachyon condensation,''
   Nucl.\ Phys.\ B {\bf 614}, 101 (2001)
   [arXiv:hep-ph/0107058]; G W Gibbons, {\it Cosmological Evolution of the
Rolling Tachyon}, \hepth{0204008}; M. Fairbairn and M.H. Tytgat,
\hepth{0204070}; T. Padmanabhan , Phys. Rev. D66, 021301 (2002),
[hep-th/0204150]; D. Choudhury, D. Ghoshal, D.P. Jatkar and S.
Panda, \hepth{0204204}; M. Sami, P. Chingangbam and T. Qureshi,
Phys. Rev. D66 (2002) 043530, \hepth{0205179};
  I.~Y.~Aref'eva, L.~V.~Joukovskaya and A.~S.~Koshelev,
  JHEP {\bf 0309}, 012 (2003)
  [arXiv:hep-th/0301137]; F. Leblond and A.W. Peet, JHEP 0304
(2003) 048,
\hepth{0303035}; S. Nojiri and S.D. Odintsov, \hepth{0306212};
M. Garousi, M. Sami and S. Tsujikawa,
Phys. Rev. D70 (2004) 043536, \hepth{0402075}.

\bibitem{sen3} A. Sen, {\it Remarks on Tachyon Driven Cosmology},
\hepth{0312153}.

\bibitem{hs1}
  H.~Singh,
  ``More on tachyon cosmology in de Sitter gravity,''
  JHEP {\bf 0601}, 071 (2006)
  [arXiv:hep-th/0505012].

\bibitem{sinha} D. Cremades, F. Quevedo and A. Sinha, {\it Warped
Inflation in Type IIB Flux Compactifications and the Open-String
Completeness Conjecture}, \hepth{0505252}.

\bibitem{hs2}
  H.~Singh,
  ``(A)symmetric tachyon rolling in de Sitter spacetime: A universe
devoid of Planck density,''
  Nucl.\ Phys.\ B {\bf 734}, 169 (2006)
  [arXiv:hep-th/0508101].

\bibitem{zamarias}
E.~Papantonopoulos, I.~Pappa and V.~Zamarias,
  JHEP {\bf 0605}, 038 (2006)
  [arXiv:hep-th/0601152].

\bibitem{anne}
M. Majumdar and A.C. Davis,
hep-th/0304226, Phys. Rev. D{\bf 69}, 103504 (2004).


\bibitem{cai}
Y-S. Piao, R-G. Cai, X-M. Zhang and Y-Z. Zhang, Phys.Rev.
D66:121301 (2002), \hepph{0207143}.

\bibitem{hsingh} H.~Singh,
  {\it ``The N-tachyon assisted inflation,''},
  arXiv:hep-th/0608032.
  %%CITATION = HEP-TH/0608032;%%

\bibitem{kofman} L. Kofman and A. Linde, {\it Problems with Tachyon
Inflation}, \hepth{0205121};  A. Linde, {\it  Inflation and string
Cosmology} \hepth{0503195}.


\bibitem{garcia}
J. Garcia-Bellido, R. Rabadan and F. Zamora, JHEp 0201
(2002) 036, \hepth{0112147};
R. Blumenhagen, B. Kors, D.
Lust and T. Ott, Nucl. Phys. {\bf B641} (2002) 235, \hepth{0202124}

\bibitem{keshav}
K. Dasgupta, C. Herdeiro, S. Hirano and R. Kallosh,
Phys. Rev. D65 (2002) 126002, \hepth{0203019};
F. Koyama, Y. Tachikawa and T. Watari, \hepth{0311191};
J. P. Hsu and R. Kallosh, JHEP 0404 92004) 042, \hepth{0402047};
 K. Dasgupta, J. P. Hsu, R. Kallosh, A. Linde and M.
Zagermann, JHEP 0408 (2004) 030, \hepth{0405247}.

\bibitem{chen}
  P.~Chen, K.~Dasgupta, K.~Narayan, M.~Shmakova and M.~Zagermann,
  %``Brane inflation, solitons and cosmological solutions: I,''
  JHEP {\bf 0509}, 009 (2005)
  [arXiv:hep-th/0501185].

\bibitem{racetreck}
J. Blanco-Pillado, C.P. Burgess, J.M. Cline, C. Escoda, M. Gomez-Reino,
R. Kallosh, A. Linde and F. Quevedo, JHEP 0411 (2004) 063,
\hepth{0406230};
B. Greene and A. Weltman, \hepth{0512135}.

\bibitem{sen1} A. Sen, {\it Rolling Tachyon}, JHEP 0204
(2002) 048, \hepth{0203211}; A. Sen, {\it Tachyon Matter}, JHEP
0207 (2002) 065, \hepth{0203265}; A. Sen, {\it Tachyon Dynamics in
Open String Theory}, \hepth{0410103}.

\bibitem{kutasov1}
  D.~Kutasov,
  {\it ``D-brane dynamics near NS5-branes,''}
  arXiv:hep-th/0405058;
  %%CITATION = HEP-TH/0405058;%%
%\cite{Kutasov:2004ct}
%\bibitem{Kutasov:2004ct}
  D.~Kutasov,
  {\it ``A geometric interpretation of the open string tachyon,''}
  arXiv:hep-th/0408073.
  %%CITATION = HEP-TH/0408073;%%

\bibitem{NST-BS} Y.~Nakayama, Y.~Sugawara and H.~Takayanagi,
JHEP  0407(2004) 020 [arXiv:hep-th/0406173]; B.~Chen, M.~Li and
B.~Sun, JHEP {\bf 0412}, 057 (2004) [arXiv:hep-th/0412022];
Y.~Nakayama, K.~L.~Panigrahi, S.~J.~Rey and H.~Takayanagi, JHEP
{\bf 0501}, 052 (2005) [arXiv:hep-th/0412038].
\bibitem{Rbrane}
K.~L.~Panigrahi, Phys. Lett. B  601 (2004) 64 (2004)
[arXiv:hep-th/0407134]; D.~A.~Sahakyan, JHEP 0410 (2004) 008
(2004) [arXiv:hep-th/0408070]; J.~Kluson, JHEP {\bf 0411}, 013
(2004) [arXiv:hep-th/0409298], JHEP {\bf 0503}, 032 (2005)
[arXiv:hep-th/0411014], JHEP {\bf 0503}, 044 (2005)
[arXiv:hep-th/0501010], JHEP {\bf 0503}, 071 (2005)
[arXiv:hep-th/0502079], [arXiv:hep-th/0504062]; O.~Saremi,
L.~Kofman and A.~W.~Peet, [arXiv:hep-th/0409092]; S.~Thomas and
J.~Ward, JHEP {\bf 0502}, 015 (2005) [arXiv:hep-th/0411130],
[arXiv:hep-th/0501192],
[arXiv:hep-th/0502228], %[arXiv:hep-th/0504226];
%[arXiv:hep-th/0508085]
B.~Chen and B.~Sun, [arXiv:hep-th/0501176]; J.~Kluson and
K.~L.~Panigrahi, JHEP {\bf 0508}, 033 (2005)
[arXiv:hep-th/0506012]; Y.~Nakayama,
  %``Black Hole - String Transition and Rolling D-brane,''
  arXiv:hep-th/0702221.
  %%CITATION = HEP-TH/0702221;%%

\bibitem{NS5}A.~Sen,
  %``Geometric tachyon to universal open string tachyon,''
  JHEP {\bf 0705}, 035 (2007)
  [arXiv:hep-th/0703157];
  %%CITATION = JHEPA,0705,035;%%
D.~Israel,
  %``Comments on geometric and universal open string tachyons near fivebranes,''
  JHEP {\bf 0704}, 085 (2007)
  [arXiv:hep-th/0703261];
  %%CITATION = JHEPA,0704,085;%%
J.~Kluson and K.~L.~Panigrahi,
  %``On the Universal Tachyon and Geometrical Tachyon,''
  JHEP {\bf 0706}, 015 (2007)
  [arXiv:0704.3013 [hep-th].
  %%CITATION = JHEPA,0706,015;%%

\bibitem{NS-cosm}H.~Yavartanoo,  [arXiv:hep-th/0407079]; A.~Ghodsi and
A.~E.~Mosaffa, Nucl.\ Phys.\ B {\bf 714}, 30 (2005)
[arXiv:hep-th/0408015]; S.~Thomas and J.~Ward,
  %``Inflation from geometrical tachyons,''
  Phys.\ Rev.\  D {\bf 72}, 083519 (2005)
  [arXiv:hep-th/0504226].
  %%CITATION = PHRVA,D72,083519;%%
%\cite{Panda:2005sg}
%\bibitem{Panda:2005sg}
  S.~Panda, M.~Sami and S.~Tsujikawa,
  %``Inflation and dark energy arising from geometrical tachyons,''
  Phys.\ Rev.\  D {\bf 73}, 023515 (2006)
  [arXiv:hep-th/0510112];
  %%CITATION = PHRVA,D73,023515;%%
%\cite{Panda:2006mw}
%\bibitem{Panda:2006mw}
  S.~Panda, M.~Sami, S.~Tsujikawa and J.~Ward,
  %``Inflation from D3-brane motion in the background of D5-branes,''
  Phys.\ Rev.\  D {\bf 73}, 083512 (2006)
  [arXiv:hep-th/0601037].
  %%CITATION = PHRVA,D73,083512;%%






\bibitem{liddle} A. Liddle, A. Mazumdar and F.E. Schunck, {\it Assisted
Inflation},
astro-ph/9804177;
A.~Jokinen and A.~Mazumdar,
   Phys.\ Lett.\ B {\bf 597}, 222 (2004)
   [arXiv:hep-th/0406074].

%\bibitem{becker}
%K. Becker, M. Becker and  A. Krause, Nucl. Phys. {\bf B175}
%(2005) 349, \hepth{0501130}; A. Ashoorioon and A. Krause,
%\hepth{0607001}.

\bibitem{sasaki} M. Sasaki and E.D. Stewart, Prog. Theor. Phys. {\bf 95}
(1996) 71.

\bibitem{Baumann}%\cite{Baumann:2007np}
%\bibitem{Baumann:2007np}
  D.~Baumann, A.~Dymarsky, I.~R.~Klebanov, L.~McAllister and P.~J.~Steinhardt,
  %``A Delicate Universe,''
  arXiv:0705.3837 [hep-th];
  D.~Baumann, A.~Dymarsky, I.~R.~Klebanov and L.~McAllister,
  %``Towards an Explicit Model of D-brane Inflation,''
  arXiv:0706.0360 [hep-th].
  %%CITATION = ARXIV:0706.0360;%%



\bibitem{wmap} See; U. Seljak, astro-ph/0407372; C.L. Bennett {\it et
al.},
Astrophys. J. Suppl, 148 (2003) 1, \hepth{0302207};
D.D. Spergel {\it et al.},
Astrophys. J. Suppl, 148 (2003)
175, \hepth{0302209}.


%\bibitem{sen0}
%A. Sen, JHEP 9910 (1999) 008, \hepth{9909062}; M.R. Garousi,
%Nucl. Phys. B584 (2000) 284, \hepth{0003122}; E.A. Bergshoeff, M. de Roo,
%T.C. de Wit, E. Eyras and S. Panda, JHEP 0005 (2000) 009, \hepth{0003221};
%J. Kluson, \hepth{0004106}; D. Kutasov and V. Niarchos, \hepth{0304045}.






%\bibitem{mazum2}
%A.~R.~Frey, A.~Mazumdar and R.~Myers,
%   {\it Stringy effects during inflation and reheating},
%   Phys.\ Rev.\ D {\bf 73}, 026003 (2006)
%   [arXiv:hep-th/0508139].
%   %%CITATION = HEP-TH 0508139;%%
%\bibitem{chialva}
%D. Chialva, G. Shiu and B. Underwood,
%{\it Warped Reheating in Multi-Throat Brane Inflation},
%JHEP 0601 (2006) 014, \hepth{0508229}.





\end{thebibliography}
\end{document}